\documentclass{aa} 
\usepackage{graphicx}
 
\begin{document}

\thesaurus{08        %
           ( 9.16.2  
             9.11.1  
             9.01.2  
                   }    

\title{Tomography of the low excitation planetary nebula NGC 40}
\author{F.\ Sabbadin\inst{1} \and  E. Cappellaro\inst{1} \and 
S. Benetti\inst{3} \and M. Turatto\inst{1} \and 
C. Zanin\inst{2,4}}
\institute{Osservatorio Astronomico di Padova, vicolo dell'Osservatorio 5,
I-35122 Padova, Italy \and Dipartimento di Astronomia, Universit\'a di
Padova, vicolo dell'Osservatorio 5, I-35122 Padova, Italy \and Telescopio 
Nazionale Galileo, Aptdo. Correos 565, E-38700 Santa Cruz de la Palma, 
Canary Island, Spain \and Institut f\"ur Astronomie der 
Leopold-Franzens-Universit\"at Innsbruck, Technikerstra\ss e 25, A-6020 
Innsbruck, Austria}

\offprints{F. Sabbadin}

\date{Received ................; accepted ................}

\maketitle

\begin{abstract}

Spatially resolved, long-slit echellograms at different position angles 
of the bright, low excitation 
planetary nebula NGC 40 indicate that, the higher is the gas excitation, the
faster is the radial motion, thus confirming the overturn of the Wilson 
law already
 suggested by Sabbadin and Hamzaoglu (1982). 

New reduction procedures, giving the radial trends of the 
electron density and of the ionic and chemical abundances, were applied to
NGC 40; they show that:

- the radial matter distribution has a sharp ``bell'' profile with peaks up
to 4000 cm$^{-3}$;

- the ionization structure is peculiar, indicating the
presence of chemical composition gradients within the nebula: the
innermost regions, hydrogen depleted, are essentially constituted 
of
photospheric material ejected at high velocity by the WC8 nucleus.

Moreover, detailed H$^+$, O$^{++}$ and N$^+$ tomographic maps,
giving the spatial ionic distributions at four position angles, are presented
and discussed within the interacting winds evolutionary model. 

\end{abstract}

\keywords{planetary nebulae: individual: NGC 40 -- ISM: kinematics and 
          dynamics -- ISM: abundances}

\section{Introduction}

\begin{figure}
\vskip 2cm
\caption{Broad-band R image of NGC 40 
(exposure time 180s, seeing 0.85$\arcsec$) obtained with the 
3.5m Italian National Telescope (TNG), revealing the intricate 
H$\alpha$+[NII] nebular structure. North is 
up and East is on the left.}\label{figura0log}
\end{figure}

NGC 40 (PNG120.0+09.8, Acker et al., 1992; Figure 1) is a very low excitation
planetary nebula powered by a WC8 star presenting a mass-loss rate of the
order of 10$^{-6}$ - 10$^{-8}$ M$_\odot$yr$^{-1}$  with wind velocities 
between 1800 and
2370 km s$^{-1}$ (Cerruti-Sola and Perinotto, 1985; Bianchi, 1992).

The nebula is described by Curtis
(1918) as a truncated ring, from the end of which extend much fainter
wisps; the brighter central portion is 38x35$\arcsec$ in PA=14$^0$, 
while the total length along this axis is about 60\arcsec. Deep, 
narrow-band imagery by Balick et al. (1992) reveals the presence
of an external network of knotty floccules and smooth filaments. Meaburn 
et al. (1996) identify two haloes around the barrel-shaped core, and  
a jet-like feature projecting from it. 

Detailed low resolution spectroscopy (Aller 
and Czyzak, 1979, 1983; Clegg et al., 1983) indicates point to point
changes of the excitation and chemical abundances which are typical of
planetary nebulae, without apparent contamination by the fast stellar
wind. 

A kinematical analysis secured by Sabbadin and
Hamzaoglu (1982) suggests the presence of peculiar motions in the central 
part of NGC 40, with the expansion velocity larger in [OIII] than in H$\alpha$ and
[NII]. 

More recently, Meaburn et al. (1996) obtained long-slit H$\alpha$
and [NII] high resolution spectra of the bright core, of the two haloes 
and of the
jet-like feature. These authors showed that the outer filamentary halo
is practically inert and that the motion of the inner, diffuse,
spherical halo mimics the one observed over the bright nebular core;
moreover, the jet-like structure is kinematically associated with the
receding end of the barrel shaped-core and does not present any of the
characteristics expected of a true jet.

In order to study in detail the nebular physical conditions and to 
apply an original procedure 
giving the spatial matter distribution along the cross-section covered by the 
spectroscopic slit,
on December 1998 we obtained a number of echellograms at different position 
angles of a dozen of winter
planetary nebulae (NGC 40, 650-1, 1501, 1535, 2022, 2371-2, 2392, 7354 and 
7662, J 320
and 900, A 12 and M 1-7). 

In this paper we report the results derived for
NGC 40. In Section 2 we present the observational material; in Section 3 we
discuss the expansion velocity field; in Section 4, the electron temperature
and the electron density are analysed and a new method to determine the radial
distribution of the electron density in expanding nebulae is presented. 
The radial ionization 
structure of NGC 40 is given in Section 5 and the resulting chemical 
composition gradients in Section 6. Section 7 shows the H$^+$, O$^{++}$ 
and N$^+$ tomographic maps derived at four position angles; a short discussion
is contained in Section 8 and conclusions are drawn in Section 9.

\begin{figure*}
\vskip 2cm
\caption{Representative examples of emission line structures observed in
NGC 40 at different position angles. Each line intensity was multiplied 
by the factor given in parenthesis to make it comparable with H$\alpha$.
}\label{fig1}
\end{figure*}

\section{Observational material}

Broad-band U, V and R imagery of NGC 40 was obtained with the Optical Imager 
Galileo (OIG) camera mounted on the Nasmyth A adapter interface (scale=5.36 
$\arcsec$ mm$^{-1}$) of the 3.5m Italian 
National Telescope (TNG, Roque de los Muchachos, La Palma, Canary Islands).
The OIG camera was equipped with a mosaic of two thinned and back-illuminated 
EEV42-80 CCDs with 2048x4096 pixels each (pixel size=13.5 microns; pixel scale=
0.072 $\arcsec$ pix$^{-1}$). 

Though these images were taken during the testing period of the instrument 
(when interference filters were unavailable), they reveal new morphological
details of the nebula supporting our spectroscopic results and will be 
presented later-on (Section 5).

Spatially resolved, long-slit spectra of the bright core of NGC 40 in the 
range $\lambda$$\lambda$4500-8000 \AA\/ (+flat fields+Th-Ar calibrations+
comparison star spectra) were obtained with the
Echelle spectrograph (Sabbadin and Hamzaoglu, 1981, 1982) attached to the
Cassegrain focus of the 182cm Asiago telescope, 
combined with a Thompson 1024x1024 pixels CCD. 

Four position angles
were selected: 20$^0$ (apparent major axis), 110$^0$ (apparent minor axis),
65$^0$ and 155$^0$ (intermediate positions). In all cases the slit-width
was 0.200 mm (2.5$\arcsec$ on the sky), corresponding to a spectral
resolution of 13.5 km s$^{-1}$ (1.5 pixel). During the exposures the
slit grazed the bright central
star (we tried to avoid most of the continuum 
contamination; the faint stellar spectrum present in our echellograms was
used as reference for the nebular centre and to correct the observations for
seeing and guiding errors). 

All spectra were calibrated in wavelength and flux in the standard way using 
the IRAF data reduction software. 

The following nebular emissions were detected: H$\beta$, 
$\lambda$4959 \AA\/  and $\lambda$5007 \AA\/  of [OIII], 
$\lambda$5755 \AA\/  of [NII], 
$\lambda$5876 \AA\/  of HeI, $\lambda$6300 \AA\/  and $\lambda$6363 \AA\/  of 
[OI] (the first at the order edge; both partially blended with night-sky 
lines), $\lambda$6548 \AA\/  and $\lambda$6584 \AA\/  of [NII], H$\alpha$, 
$\lambda$6578 \AA\/  of CII, 
$\lambda$6717 \AA\/ and $\lambda$6731 \AA\/  of [SII], $\lambda$7135 \AA\/  of 
[ArIII],
$\lambda$7320 \AA\/ and $\lambda$7330 \AA\/  of [OII] (at the extreme edge of 
the 
order; out of focus) and $\lambda$3726 \AA\/ and $\lambda$3729 \AA\/  of [OII] 
(second order). 

All line 
intensities (but not the second order doublet of [OII]) were de-reddened 
by fitting the observed H$\alpha$/H$\beta$ ratio to the one
computed by Brocklehurst (1971) for Te=$10^4$ K and Ne=$10^4$
cm$^{-3}$. We derive a logarithmic extinction at H$\beta$, c(H$\beta$)=$0.60
\pm 0.10$, to be compared with the values of 0.28, 0.65 and 0.70
obtained by Kaler (1976), Aller and Czyzak (1979) and Clegg et
al. (1983), respectively.
The fair agreement (within 10\%) between our integrated line
intensities and those reported in the literature from low resolution
spectra (see, for example, Clegg et al., 1983) induced us to adopt 
 I(3726+3729)/I(H$\alpha$)= 1.48, as reported by these
authors.  

A large variety of emission structures is present in NGC 40;
for illustrative purposes, some examples are given in
Figure 2. In these reproductions the faint stellar 
continuum was
removed and the observed intensities enhanced (by the factor
given in parenthesis) to make each line comparable with
H$\alpha$.  

Our spectra at PA=110$^0$
(apparent minor axis, Fig. 2a) were centred 2$\arcsec$ South of the 
central
star. All emissions present the classical bowed shape expected for
 expanding shells of different mean radii $r$ and thicknesses
$t$, ranging from $r=16''$  for [OIII] and CII to $r=19''$  for
[NII] and [SII] and from $t=0.20r$ for [NII] and [SII] to $t=0.25r$ for 
[OIII]
and CII (as also suggested by the integral intensity
distributions along the slit). 

At PA=20$^0$ (apparent major
axis; slit centre 2$\arcsec$ West of the star; Fig. 2b) low excitation
lines (such as the [SII] red doublet) show knotty structures and fail to 
close at either  polar
cap. High excitation emissions (for instance [OIII]) appear more
homogeneous; they merge in a diffuse zone internal to the polar caps;
the smooth intensity distribution and the large velocity spread 
suggest that [OIII] completely fills this part of nebula;
moreover, the faintness (or absence) of an external low excitation
counterpart indicates that in these directions the main body of NGC 40
is density bounded rather than ionization bounded. 

Spectra taken al
PA=65$^0$ (Fig. 2c) were centred 2$\arcsec$  S-E of the central star; 
they appear quite regular, but for a deformation of the blue-shifted
component in the E-NE quadrant. 

At PA=155$^0$ (centre of the slit 
2$\arcsec$ S-W of the star; Fig. 2d) low excitation lines (such as the 
[NII] line at $\lambda$6584 \AA\/) merge in the N-NW sector, 
in correspondence of a bright knot. The S-SE region is strongly 
depauperated of gas and only high excitation emissions (like the extremely 
faint CII line at $\lambda$6578 \AA\/) present a close structure; as already 
seen for PA=20$^0$, the weakness of external low excitation emissions 
suggests that in these directions the nebula is density bounded.

\begin{figure}
\resizebox{\hsize}{!}{\includegraphics*{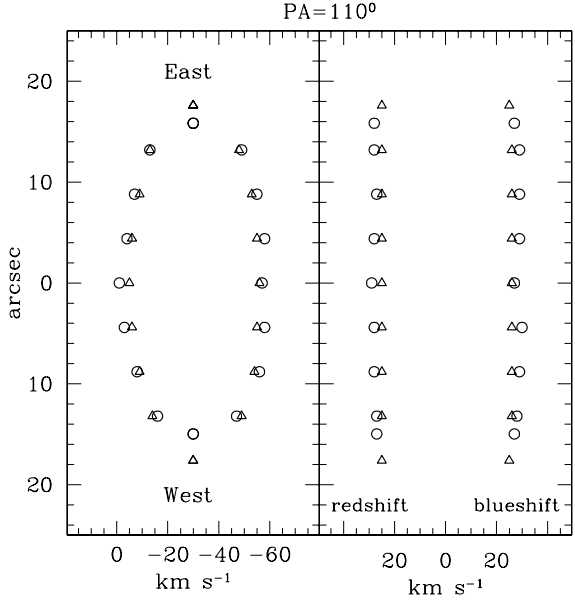}}
\caption{Observed (left panel) and deprojected (right panel) expansion
velocity field at PA=110$^0$ (apparent minor axis). Triangles are for [NII] 
and circles for [OIII].}\label{fig2}
\end{figure}

\begin{figure}
\resizebox{\hsize}{!}{\includegraphics*{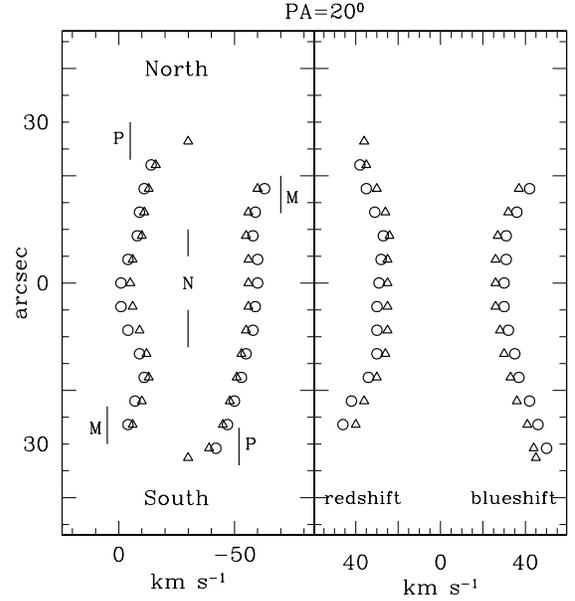}}
\caption{Same as Fig. 3, but for PA=20$^0$ (apparent major axis). 
Glossary: N=main nebula,
P=polar caps (blue-shifted at south, red-shifted at north), M=extremely
faint H$\alpha$ and [NII] mustaches (red-shifted at south, blue-shifted
at north) emerging from the main nebular body.}\label{fig3}
\end{figure}

\section{The expansion velocity field}

First of all we measured the peak separation at the centre of each
nebular emission. Results are presented in Table 1, where ions are put in order
of increasing ionization potential.

\begin{table}
\caption{Expansion velocities}
\begin{flushleft}
\begin{tabular}{lcrcc}
\hline\noalign{\smallskip}
~~$\lambda$  & ion &  I.P.  &      $2\,V_{exp}$ & notes\\
~(\AA)      &     &   (eV) &      (km s$^{-1}$) \\
\noalign{\smallskip}
\hline\noalign{\smallskip}
6300&       [OI]&     0   &     51 $\pm$2&         1\\
6717-6731&  [SII]&    10.4&     51 $\pm$2& \\
6563&        HI&      13.6&     50 $\pm$1&\\
3726-3729&  [OII]&    13.6&     54 $\pm$2&         2\\
6584&       [NII]&    14.5&     51 $\pm$1&\\
6578&        CII&     24.4&     53 $\pm$2&\\
5876&        HeI&     24.6&     53 $\pm$2&\\
7135&       [ArIII]&  27.6&     54 $\pm$2&  \\   
5007&       [OIII]&   35.1&     57 $\pm$1&\\
\noalign{\smallskip}
\hline
\end{tabular}
\end{flushleft}

1 = partially blended with night-sky emission\\
2 = second order lines
\end{table}

We find a complete overturn of the Wilson law: in NGC 40 the higher is the
ionization, the faster is the motion (thus confirming the suggestion by 
Sabbadin and Hamzaoglu, 1982, of a peculiar [OIII] expansion velocity). 

A further peculiarity is represented by [OII], which expands faster than 
the other low ionization 
species. This [OII] velocity-excess is related to the very low degree
of excitation of the nebula: since the O$^+$ zone stops at 35.1 eV
(while, for example, S$^+$ falls at 23.4 eV and N$^+$ at 29.6 eV), it extends 
well
inwards, as also indicated by the [OIII] weakness. So, in NGC 40, [OII] must 
be considered a medium-high excitation species (at least for the dynamical 
point of view). 

Clearly, the expansion velocities reported in Table 1 are only
mean values and do not describe the complex kinematics of the nebula. 
To better illustrate this, we have selected two position
angles (110$^0$=apparent minor axis; 20$^0$=apparent major axis) and chosen 
[NII] for low excitation regions and [OIII] for the high excitation ones 
("high" at least for NGC 40!). 

The regular ellipses shown by these ions 
at PA=110$^0$ (Figure 3, left panel) are typical of expanding shells; 
the absence of 
tilt indicates that the nebular cross-sections are either 
circular, or elliptical (in this second case we are aligned with one of the 
axes). In fact,
the de-projection of the observed expansion velocities (assuming a simple,
direct position-speed correlation) gives the "true" expansion velocities 
presented in the right panel of Figure 3, i.e. a constant value for each ion. 
Similar trends
are obtained also at PA=65$^0$ and in the N-NW sector of PA=155$^0$ .

The [NII] and [OIII] expansion velocity fields observed at PA=20$^0$ (major
axis) are given in Figure 4 (left panel). 
Here the 
situation is more complicated, mainly with respect to [OIII]. In fact, 
while [NII] 
presents sharp emissions along the slit, [OIII] describes an ellipse which 
is well defined in the main part of the nebula, but becomes smooth and broad 
to the north and south. The de-projection of the observed velocities 
(right panel 
of Figure 4) 
points out the acceleration suffered by the polar (less dense) 
nebular material. 
A similar behavior occurs at PA=155$^0$, in the S-SE sector.  

For a better understanding of the peculiar kinematics of NGC 40, 
a detailed analysis of the nebular physical conditions can be useful.

\begin{figure}
\resizebox{\hsize}{!}{\includegraphics*{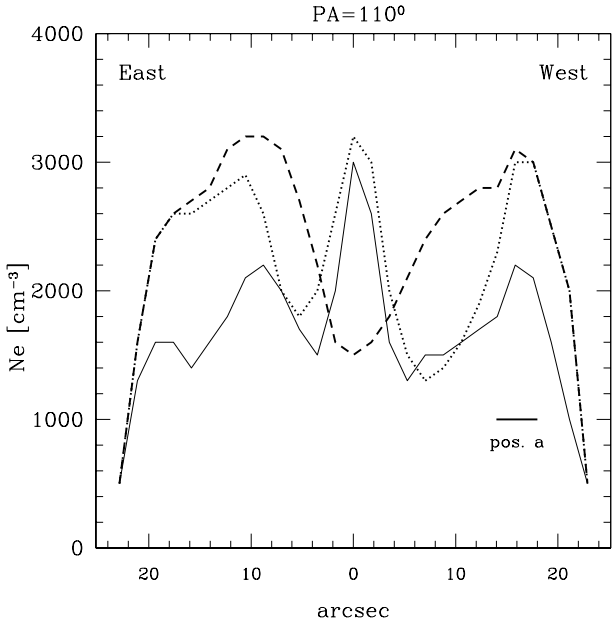}}
\caption{[SII] electron density trends observed at PA=110$^0$ for the
blue-shifted (dotted curve) and red-shifted (dashed curve) peaks and
for the integrated spectrum (solid curve). ``Position a'' of Clegg et al.
(1983) is indicated.}\label{fig4}
\end{figure}

\begin{figure}
\resizebox{\hsize}{!}{\includegraphics*{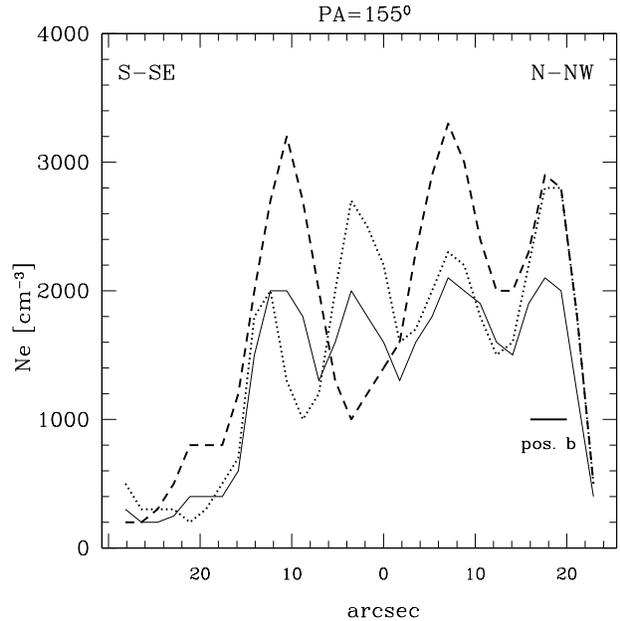}}
\caption{Same as Fig. 5, but for PA=155$^0$. ``Position b'' of Clegg et
al. (1983) is indicated.}\label{fig5}
\end{figure}

\section{Electron temperature (Te) and electron density (Ne)}

The only $Te$ diagnostic present in our spectra is the 
line intensity ratio 6584/5755 \AA\/ of [NII]. 
Unfortunately, the auroral 
emission is 
too weak for a detailed analysis; for the brightest parts of NGC 40 we 
obtain I(6584)/I(5755)=125$\pm$20, corresponding to $Te$=7900 
$\pm$200 K (assuming Ne=10$^4$ cm$^{-3}$). Thus, according to Aller 
and Czyzak (1979, 1983) and Clegg et al. (1983), we 
will adopt $Te$=8000 K for low ionization regions (O$^0$, N$^+$, S$^+$, 
O$^+$) 
and $Te$=10000 K for the others (O$^{++}$, He$^+$, C$^+$, Ar$^{++}$).  

Our $Ne$ study is essentially based on [SII] lines at $\lambda$6717   \AA\/
 and 
$\lambda$6731 \AA. 

To test the reliability of our measurements we 
can make a comparison with the results of Clegg et al. (1983), 
who observed at low spectral resolution a patch of nebulosity located at 
north-west of the central star. At "position a" they found 
I(6717)/I(6731)=0.68, corresponding to $Ne$=2200 cm$^{-3}$ (for $Te$=8000 K). 
"Position a" practically coincides with the bright western edge 
of our spectra taken at PA=110$^0$ (apparent minor axis). 

The [SII] 
density distribution we observed at this position angle is shown in Figure 
5 for the blue-shifted and the red-shifted peaks, and also for the integrated, 
low resolution spectrum (we have averaged our echellograms 
over a velocity range of 312 km s$^{-1}$, which corresponds to the 
spectral resolution used by Clegg et al., 1983). The estimated $Ne$ accuracy
varies from $\pm$15\% for high density peaks (generally -but not always-
coinciding with the strongest emissions), to $\pm$50\% for deep valleys
(weakest components). 

For "position a" the
integrated spectrum gives $Ne=$2100 cm$^{-3}$, in good agreement with Clegg 
et al. (1983), but high resolution measurements indicate a density of
3100 cm$^{-3}$. Note that in Figure 5 the maximum density of the
integrated spectrum (3000 cm$^{-3}$) is reached in a relatively weak knot
projected onto the central star; it corresponds to a peak (3200
cm$^{-3}$) in the blue-shifted component. 

As a further example, in Figure 6
we present the [SII] density distribution observed at PA=155$^0$;
symbols are as in Figure 5. In this case our N-NW peak is
partially superimposed to ``position b'' of Clegg et al. (1983). 
Once again their electron density ($Ne$=2100 cm$^{-3}$)
coincides with our integrated value, but is considerably lower than
the peak measured at high resolution ($Ne$=2900 cm$^{-3}$).

{\bf CAVEAT}:{\em  strictly speaking, our Ne estimates too must be
considered as lower limits, due to ionization effects. In fact, as we
will see in the next section, the complete recombination of
S$^{++}$ occurs externally to the top of the density distribution;
this implies that the peak of the [SII] intensity distribution is slightly
outwards with respect to the density maximum, leading to a 
small underestimate of Ne derived from the [SII] line intensity
ratio.}

\begin{figure}
\resizebox{\hsize}{!}{\includegraphics*{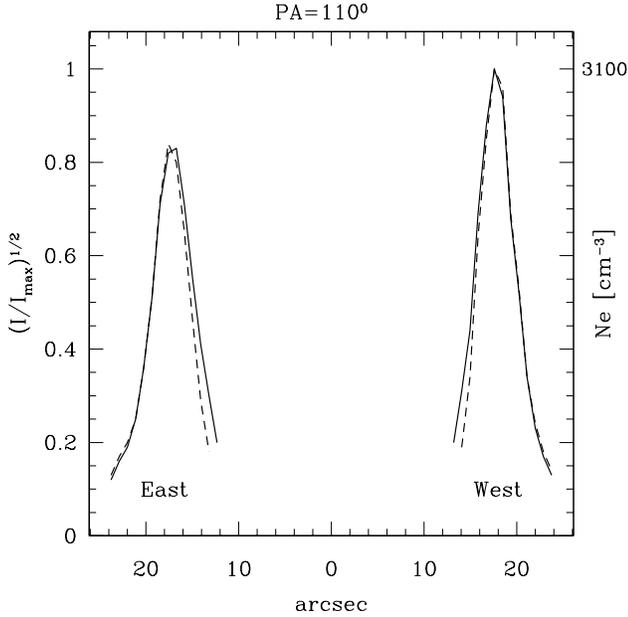}}
\caption{Relative (left ordinate scale) and absolute (right ordinate scale) 
radial electron
density distributions at PA=110$^0$ from the H$\alpha$ (solid
curve) and [NII] (dashed curve) zero-velocity columns.}\label{fig6}
\end{figure}

\begin{figure}
\resizebox{\hsize}{!}{\includegraphics*{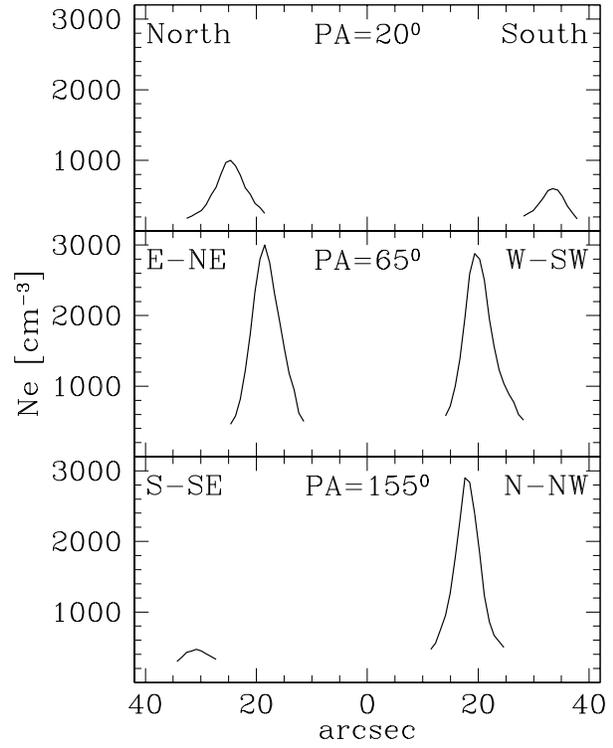}}
\caption{Radial electron density profiles obtained from the H$\alpha$ 
zero-velocity 
columns at PA=20$^0$, 65$^0$ and 155$^0$.}\label{fig7}
\end{figure}

Besides the density peak along the cross-section of the nebula, our high 
resolution spectra allow to derive the radial distribution of the
electron density using the
"zero-velocity column" of each emission line; this is the central,
rest-frame pixel column containing the ionized gas moving in the 
neighborhood of -30 km
s$^{-1}$ (mean radial velocity of the whole nebula). In practice, 
the zero-velocity
column contains the material which is expanding normally to the line 
of sight; it is important because it insulates a definite slide of nebula 
which is unaffected by the expansion velocity field. 

The intensity distribution observed in the zero-velocity column of each 
nebular emission
must be 
cleaned of the broadening effects due to:

- instrumental profile, corresponding to a Gaussian having FWHM=13.5 
km s$^{-1}$;

- thermal motions; for $Te$=8000 K they amount to 19 km s$^{-1}$ for H, 
9.5 km s$^{-1}$ for He and 5 km s$^{-1}$ (or less) for the heavier elements;

- turbulent motions, which affect all lines in the same way; following Meaburn
et al. (1996) in NGC 40 they are 6 - 8 km s$^{-1}$;

- fine structure of $H\alpha$ and $\lambda$5876 \AA\/ of HeI. Following Dyson 
and
Meaburn (1971) and Dopita (1972), the seven H$\alpha$ components can be
shaped
as the sum of two equal Gaussian profiles separated by 0.14 \AA\/. 
$\lambda$5876 \AA\/ of HeI has six components: five of them are concentrated 
in the
narrow range $\lambda$$\lambda$5875.5987 - 5875.6403 \AA\/ (they were 
considered a single emission at $\lambda$5875.62 \AA\/); the sixth one, 
located at
$\lambda$5875.9663 \AA\/, is 4.4 times weaker (and then it was neglected). 

We have derived the corrected intensity of the zero-velocity columns by 
deconvolving the 
observed
profile at any given nebular position along the slit into a series 
of Gaussians 
spaced 
of one pixel along the dispersion and having: FWHM=25.8 km s$^{-1}$ for 
H$\alpha$, 17.9 km s$^{-1}$ 
for $\lambda$5876 $\AA$ of HeI and FWHM=16.0 km s$^{-1}$ for the other lines.
We have cautiously 
stopped the computation when
I$_{obs}$$\gid$2I$_{corr}$, corresponding to an uncertainty in $Ne$ of
$\pm$30\%. 

The last
step is to deconvolve the corrected zero-velocity columns for guiding errors 
and 
seeing
by means of the profile, along the slit, of the
stellar continuum present in the spectrum.  

At each position
of the zero-velocity column the intensity is proportional to $NeNi$ and,
in the case of total ionization and Te=constant, I$\propto$$Ne$$^2$. This
is valid, separately, for all ions. 

As an example, the
H$\alpha$ and $\lambda$6584 \AA\/ of [NII] zero-velocity columns
derived at PA=110$^0$ (apparent minor axis) are presented in
Figure 7; densities are normalized to the strongest
peak (left ordinate scale).  As expected, $Ne$(H$^+$) and $Ne$(N$^+$) 
practically
coincide in the external parts (up to the maxima), and tend to diverge
moving inwards, due to the gradual double ionization of nitrogen. Let's
consider the H$^+$ curve; if we scale the western peak to $Ne$=3100
cm$^{-3}$, as previously derived from the [SII] line intensity ratio
(see Fig. 5), we obtain the true density distribution along the
central, zero-velocity column (right ordinate scale).To be noticed that the 
density of the eastern peak results to be 2600 cm$^{-3}$,
in fairly good agreement with the value derived from [SII]
diagnostics.  

In the same way we obtained the radial density trends at the
other position angles; they are given in Figure 8. Note that the
profiles observed at PA=20$^0$ correspond to the northern ($Ne$=1000
cm$^{-3}$) and southern ($Ne$=600 cm$^{-3}$) polar caps, and that only
a weak emission ($Ne$=400 cm$^{-3}$) is visible in the S-SE
sector of PA=155$^0$. 

In the end, the "bell" profile seems to be a
general characteristic of the radial matter distribution within NGC 40.  

A
final note on the $Ne$ problem in this nebula concerns [OII]. We have
repeated the previous procedures using the 3726/3729 diagnostic ratio;
the [OII] density trends are, in all cases, very close to the [SII]
ones, but the actual density is systematically lower by a factor of almost 
two.  To
explain this curious feature, we must examine in detail the ionization
structure of NGC 40.

\begin{figure}
\resizebox{\hsize}{!}{\includegraphics*{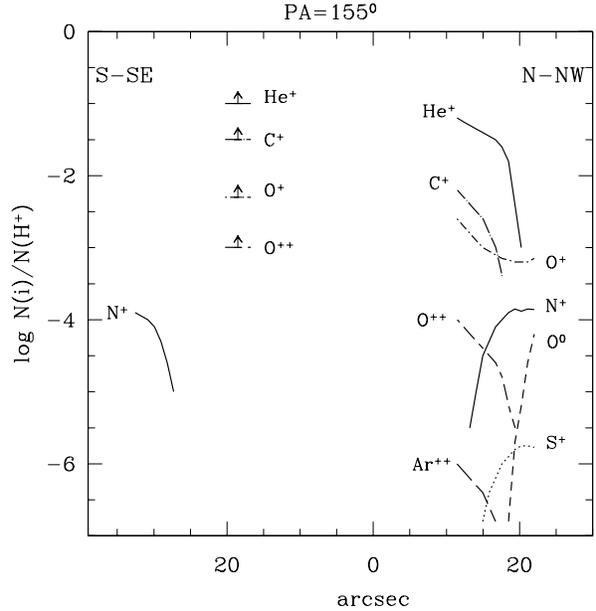}}
\caption{Radial ionization structure of NGC 40 at PA=155$^0$ obtained from 
the zero-velocity columns. Details are given in the text.}\label{fig8}
\end{figure}

\section{Radial ionization structure}

The detailed radial ionization structure of an expanding nebula can be 
derived by comparing the zero-velocity columns of different lines. 

To obtain the ionic abundances (relative to 
H$^+$) of O$^0$, O$^+$, O$^{++}$, N$^+$, C$^+$, S$^+$, He$^+$ and Ar$^{++}$ 
we have applied the classical procedure (see, for instance, Peimbert and 
Torres-Peimbert, 1971; Barker, 1978; 
Aller and Czyzak, 1979, 1983). 

For all ions, but not [ArIII], we have used the 
same atomic 
constants adopted by Clegg et al. (1983) in their low resolution study of 
NGC 40; for [ArIII], not mentioned by these authors, transition probabilities
were taken from Mendoza (1983) and collision strengths from Galavis et al.
(1995). 

Since the H$^+$ central column is practically absent in
the S-SE sector of PA=155$^O$ (while other ions are present), only lower 
limits to the
relative ionic abundances can be obtained; the same situation occurs at
PA=20$^0$ (major axis), internally to the polar caps. 

In all the remaining 
cases (where the radial ionization structure of the nebula can be studied in 
detail) most ions present the expected trends: medium-high ionization species 
(O$^{++}$, Ar$^{++}$, He$^+$) rapidly decrease outwards; the opposite for low 
ionization species, such as O$^0$, N$^+$ and S$^+$. Once again, O$^+$ shows 
an anomalous behavior (also C$^+$ has similar characteristics, but the CII
$\lambda$6578 \AA\/  line is so weak that a quantitative analysis is 
hazardous). 

The radial ionization structure of NGC 40 is synthesized in Figure 9 for 
PA=155$^0$; this position angle is a perfect compendium of the different 
situations occurring in the nebula, since it presents a well defined H$^+$ 
structure in the N-NW part and practically no hydrogen emission at S-SE
(see Figure 8). Moreover, its N-NW peak corresponds to "position b" of 
Clegg et al. (1983), so that a direct comparison with their results can be 
made. 

Let's first consider the N-NW sector of Figure 9; here both O$^+$/H$^+$ and 
O$^{++}$/H$^+$ increase inwards (while we would expect opposite trends,
since the O$^+$$\getsto$O$^{++}$ equilibrium shifts on the left going 
outwards, 
so that when [OII] increases [OIII] falls, and vice versa). Note also that 
the O$^{++}$ curve lies well below the O$^+$ one, indicating that the 
stellar flux is unable to completely ionize O$^+$. Ionization structures 
similar (qualitatively and quantitatively) to the one observed in the N-NW 
sector of PA=155$^0$ were obtained at PA=65$^0$ and 110$^0$.

Look now at the left part of Figure 9. In the S-SE sector of PA=155$^0$ 
the H$^+$ zero-velocity column shows only a weak emission at about 
30$\arcsec$
from the center (and here the N$^+$/H$^+$ trend was derived), 
whereas high ionization species are present also at intermediate positions, 
giving the lower limits to the He$^+$/H$^+$, C$^+$/H$^+$, O$^+$/H$^+$ and 
O$^{++}$/H$^+$ abundances indicated in the figure. The degree of 
excitation in these hydrogen-depleted zones remains low, being 
N(O$^+$)/N(O$^{++}$)$\ga$5. Similar low density, hydrogen-depleted
regions are present at PA=20$^0$ (major axis), internally to the polar caps.

An independent confirmation of the existence of a peculiar 
ionization structure in NGC 40 comes from the broad-band U, V and R imagery
obtained with the 3.5m Italian National Telescope (TNG).

The R band (Figure 10 upper left) is dominated by H$\alpha$ and [NII], 
with a small
contamination of [SII] at $\lambda$6717 \AA\/ and $\lambda$6731 \AA\/.

Over 90\%\ of the U light (Figure 10 upper right) comes from the [OII] 
doublet at 
$\lambda$3726 \AA\/ and $\lambda$3729 \AA\/; the remaining is due to high
Balmer lines.

Finally, in the V band (Figure 10 lower left) H$\beta$ contributes for 55\%\ 
and 
the [OIII] doublet at $\lambda$4959 \AA\/ and $\lambda$5007 \AA\/ for 45\%\/.

Since the H$\beta$ morphology of this low excitation planetary nebula is very 
similar to 
the H$\alpha$+[NII] one, 
we can extract the [OIII] component of the V image by   
subtracting  
the R image (appropriately scaled in intensity) to the V image.
The resulting [OIII] appearance of NGC 40, shown in Figure 10 lower right,
noticeably differs from the H$\alpha$+[NII] one; this morphological 
peculiarity,
known since Louise (1981), stands out in great detail (greater than 
before; see, for instance, Balick, 1987 and Meaburn et al., 1996).

The apparent [OII]/[OIII] distribution over the nebula can be obtained
from the ratio U/[OIII]  and the(H$\alpha$+[NII])/[OII]
distribution from R/U; results are shown in Figures 11.

The large stratification effects present in these last maps indicate that 
all the three intensity ratios [OII]/[OIII],
(H$\alpha$+[NII])/[OII] and (H$\alpha$+[NII])/[OIII] decrease inwards, i.e.
all the three ionic ratios O$^+$/O$^{++}$, (H$^+$+N$^+$)/O$^+$ and 
(H$^+$+N$^+$)/O$^{++}$ decrease inwards. In conclusion: all the three ionic 
ratios
O$^{++}$/O$^+$, O$^+$/H$^+$ and O$^{++}$/H$^+$ increase inwards (being 
the H$^+$ distribution very similar to the N$^+$ one).
This is in perfect qualitative agreement with the results obtained from 
high resolution spectrography.

\begin{figure}
\vskip 2cm
\caption{Broad-band imagery of NGC 40 (seeing 0.85'') obtained with the 
3.5m Italian National Telescope (TNG). The R image (upper left, exposure time 
180s) is dominated 
by H$\alpha$ and [NII] and the U image (upper right, 600s) by the [OII] 
doublet at $\lambda$3726 and $\lambda$3729 \AA\/. The V image (lower left, 
180s) is the sum of H$\beta$ (55\%) and [OIII] at $\lambda$4959 and 
$\lambda$5007 \AA\/ (45\%). The lower right panel shows the [OIII] 
appearance of NGC 40; it was obtained by subtracting the R image 
(appropriately scaled in intensity) to the V image.}\label{colori4}
\end{figure}

\begin{figure}
\vskip 2cm
\caption{Left panel: apparent [OII]/[OIII] distribution in NGC 40. [OIII] 
is weak in 
the outermost strata (white), it gradually increases inwards, reaching its
maximum (black) internally to the polar caps.
Right panel: apparent (H$\alpha$+[NII])/[OII] distribution in NGC 40. This 
intensity ratio presents a maximum (white) in the external, bright shell, it 
decreases inwards, and has a minimum 
(black) internally to the polar caps. Same 
orientation and scale as Fig. 10.}\label{Fig. 10}
\end{figure}

The obvious conclusions emerging from the peculiarities in the
ionization structure of NGC 40 are drawn in the following, short section.

\section{Abundance gradients towards NGC 40.}

We believe that the radial ionization structure of NGC 40 indicates the 
presence 
of abundance 
gradients across the nebula (although we cannot exclude a minor effect 
due to electron temperature variations). 
Take oxygen, for instance. 

The general expression O=O$^0$+O$^+$+O$^{++}$+etc.  in the case of NGC 40 
reduces to O=O$^+$, 
i.e. the observed O$^+$/H$^+$ ratio is a good match of 
the total oxygen abundance. 
We obtain O/H=6$\pm$1 x10$^{-4}$ in the main, denser regions (to be 
compared with the values of 6.0x10$^{-4}$ and 8.4x10$^{-4}$ given 
by Aller and Czyzak, 1983, and Clegg et al., 1983, respectively). O/H rapidly
increases inwards; at the observational limit it is larger than 5x10$^{-3}$. 

Less accurate abundance gradients can be derived in a similar manner for 
C/H (1x10$^{-3}$ to $>$3x10$^{-2}$) and for He/H (4x10$^{-2}$ 
to $>$1x10$^{-1}$).

It is very tempting to attribute these abundance trends to contamination
of the stellar wind (we remember that detailed analyses of the optical
spectrum of the WC8 nucleus carried out by Leuenhagen et al., 1996,
give an upper limit for the mass fraction of hydrogen of 2\%; moreover,
He/C=0.8 and C/O=5). In this scenario, our internal, low density, 
hydrogen-depleted
regions are essentially constituted of enriched material recently
emitted by the central star. Moving outwards, this gas gradually
mixes with the shell's gas and abundances rapidly fall to the nebular
values.  

The general expression I$\propto$$NeNi$ (valid in the zero
velocity column) allows to estimate the electron density in the
innermost, fast moving, hydrogen-depleted regions. An indicative value
of 1 - 2 cm$^{-3}$ is obtained, corresponding to a few 10$^{-24}$ g
cm$^{-3}$, and (for a distance of 1100 pc) to a matter flux of a few
10$^{-8}$ M$_\odot$ yr$^{-1}$. This must be the order of magnitude of
the present central star's mass-loss.

\section{Tomography}

In order to obtain the spatial 
matter distribution within a planetary nebula we 
will analyse the observed emission line structures by means of an 
iterative procedure which is the implementation of the 
method 
originally 
proposed by Sabbadin et al. (1985, 1987 and references therein) for plate
echellograms. In this approach, the relative density distribution of an 
emitting region along the cross-section of nebula covered by the slit is 
obtainable from the radial velocity, FWHM and intensity profile. 

The calibration to absolute densities is straightforward if [SII] or [OII] 
doublets 
are present (as in the case of NGC 40); for density bounded planetary nebulae 
other diagnostics (such as 5517/5537 of [ClIII] or 4711/4740 of [ArIV]) can 
be used but, due to the intrinsic weakness of these lines, results are 
often uncertain; so, for high excitation planetary nebulae, we prefer to adopt 
the surface brightness method (it  will be described in a forthcoming paper 
dedicated to NGC 1501).

The tomographic analysis contains a large amount of physical informations;
for reasons of space too, this first application to NGC 40 intends to
be a concise test of reliability (i.e. a semi-quantitative
comparison with the observational results independently accumulated in the 
previous sections).

Amongst the different maps (intensity, density, fractional 
ionization, ionic abundance etc.) obtainable with our reconstruction method, 
we have selected 
$\sqrt{NeNi}$; H$^+$ was chosen as ionization marker, and N$^+$ 
and O$^{++}$
to symbolize low and high excitation zones, respectively. 

Since for H$^+$
we have that $\sqrt{NeNi}$ is proportional to Ne, the H$^+$ maps 
give the
relative density distributions along the cross-sections of NGC 40 covered 
by the
slit; this is roughly true also for N$^+$ (its
ionization occurring only in the innermost parts), but not for O$^{++}$, due
to the incomplete O$^+$ ionization. 

The gray-scale $\sqrt{NeNi}$ maps derived for H$^+$, O$^{++}$ and N$^+$ at 
PA=110$^o$, 20$^o$, 65$^o$ and 155$^o$ are shown in Figure 12.

At PA=110$^o$ (apparent minor axis) H$^+$ describes a circular, knotty
ring; N$^+$ closely mimics H$^+$ (but, obviously, it is a bit sharper),
while O$^{++}$ is emitted in the internal regions and rapidly drops when
$Ne$ raises. To be noticed the opposite behaviors of H$^+$ and O$^{++}$, 
in 
the sense that bright H$^+$ zones tend to be associated with faint 
O$^{++}$ regions, and vice versa; the patch studied by Clegg et al. (1983, 
their ``position a'') corresponds to our western edge.

At PA=20$^o$ (apparent major axis) the H$^+$ and N$^+$ ring structures 
appear disrupted at north and south, where the two polar caps are formed;
unfortunately, in these reproductions we have lost the two extremely 
faint (and fast moving) low excitation mustaches attached to the main nebular
body (blue-shifted at north, red-shifted at south). The elliptical structure
shown at this position angle by O$^{++}$ testifies the existence of
hydrogen-depleted zones internally to the polar caps (which, in our 
opinion, are essentially constituted of stellar wind).

The $Ne$ map derived from H$^+$ at PA=65$^o$ describes a distorted ring 
presenting a deformation in the approaching gas of the E-NE quadrant;
the bright knots visible in both the H$^+$ and N$^+$ maps completely 
disappear in the
O$^{++}$ one, which presents a more homogeneous distribution.

Finally, at PA=155$^o$ the ring structure shown by H$^+$ and N$^+$ in the N-NW
sector is completely absent at S-SE; once again, O$^{++}$ indicates the 
presence
of hydrogen-depleted gas within the S-SE region; the N-NW border corresponds
to ``position b'' of Clegg et al. (1983).   

\begin{figure*}
\vskip 2cm
\caption{Gray-scale $\sqrt{NeNi}$ spatial maps of NGC 40 in H$^+$ (left), 
O$^{++}$ 
(center) and N$^+$ (right) at different position angles. For H$^+$ we have 
that $\sqrt{NeNi}$ is proportional to Ne; this is approximately valid also 
for N$^+$, due to the very low degree of excitation of the nebula, but 
not for O$^{++}$, since the O$^+$ ionization is incomplete .}\label{tom110}
\end{figure*}

\begin{figure*}
\vskip 2cm
\caption{Isophotal density maps of NGC 40 obtained from H$^+$. Contour
levels: dashed curve=250 cm$^{-3}$, solid curves=500, 1000, 1500, 2000,
2500, 3000 and 3500 cm$^{-3}$. Same orientation as Fig. 12. In each map 
the central 
star is located 2$\arcsec$ over or below the plane of the figure.}\label
{contorni}
\end{figure*}

Quantitative maps, giving the true electron density distribution at the four 
position angles, can be obtained by calibrating the H$^+$ maps of Fig. 12 
with the [SII] line intensity ratio. Results are shown in Figure 13, where  
the 
nebular centre corresponds to the position of the exciting star (which,
actually, lies 2$\arcsec$ over or below the plane of the figure).

\section{Discussion}

The general rule given by Olin Wilson (1950) in his classical work on the
expansion velocity of planetary nebulae -"high-excitation particles show
smaller separation, and low-excitation particles higher separation"- is 
definitely violated in NGC 40, which presents a positive inwards velocity
gradient. Moreover, this velocity gradient is small enough 
($\Delta V\simeq  3 - 4$ 
km s$^{-1}$ for the main shell) to rouse the suspicion that 
some nebular lines can be produced by resonance scattering. 

The problem 
has been analysed in detail by Clegg et al. (1983), who tried to explain 
the abnormal CIV $\lambda$1549 \AA\/  intensity observed in NGC 40 in 
terms of stellar light scattered by the nebula. Assuming a spread in 
velocity of 20 km s$^{-1}$, these authors obtained that the stellar flux 
at $\lambda$1549 \AA\/ is nine times larger than the $\lambda$1549 \AA\/ 
nebular scattered flux and concluded that 
scattering is inadequate, but pointed out that "this mechanism will be most 
effective 
if velocity gradients are small". Decidedly, our new $\Delta V$ value re-opens 
the question. 

We have seen that the peculiarities found in the ionization 
structure strongly suggest the presence of abundance gradients within NGC 40; 
the responsible of these gradients is the fast stellar wind. This is not a 
surprise since, following Bianchi and Grewing (1987), the mass of the central 
star is close to the C-O core of the progenitor star. 

It is interesting 
to note the discrepancy found by these authors between the stellar temperature 
determined from IUE observations (90000 K) and the much lower value of 
30000 K inferred from the nebular ionization. They overcame the 
impasse with a carbon curtain screening the high energy photons; this 
2 - 5x10$^{18}$ cm$^{-2}$ CII column density should be located at the inner 
edge of the shell, where the fast wind from the nucleus interacts 
with the nebular gas. 

The hypothesis is suggestive, but some doubts remain, 
essentially for lack of space. In fact, on one side our ionization maps 
indicate that the CII layer producing the requested column density must 
have a thickness of the order of 10$^{19}$ cm, on the other side the mean 
nebular radius of NGC 40 (assuming a distance of 1100 pc, centre of the 
large number of individual and statistical distances reported in the 
literature) is 0.08 pc, i.e. 40 times shorter than the CII layer. An
even more unfavorable situation occurs for a nebular distance of 980 pc,
as adopted by Bianchi and Grewing (1987).

We wish to stress here the great opportunities opened by the analysis of the
zero-velocity columns in expanding nebulae (planetary nebulae, shells around 
Population I W-R stars, supernova remnants etc.): for the first time we can 
insulate a definite slide of nebula, thus removing the tiresome projective 
problems related to these extended objects.

To be noticed the strict analogy between the zero-velocity column obtained from
 slit spectroscopy and the rest-frame image derived from Fabry-Perot 
interferometry; in a certain sense, the rest-frame image is a zero-velocity 
column extended at all position angles (and this represents a considerable 
observational advantage).

Moreover, the expansion velocity of a few objects (for instance the Crab 
nebula) is so large that the ``zero-velocity column'' analysis can be applied 
to very low resolution data (slit spectroscopy or Fabry-Perot interferometry).

Concerning our tomographic results, we confess a genuine satisfaction for
the quality of these "preliminary" maps (clearly, the final goal of each
tomographic study is a complete, three-dimensional model, but this will need 
a more detailed spectroscopic coverage). 

Our figures confirm that the main
body of NGC 40 has an inhomogeneous, elongated barrel-shaped structure ("an
opened-up ellipse" following Mellema, 1995; "a slightly tilted cylindrical
sleeve" following Balick et al., 1987), with thin arcs emerging at both ends
of the major axis. 

This morphology is normally interpreted as the result of
the interaction of the fast stellar wind from the high temperature central
star with the slow, inhomogeneous (i.e. denser in the equatorial plane) 
super-wind from the AGB progenitor (see Balick et al., 1987, Mellema, 1995 
and references therein). 

Following Balick et al. (1987) the stellar wind 
creates in NGC 40 an interior bubble at high temperature and ionization. 
The hot gas cools by expansion or radiative losses in forbidden lines when it 
interacts with the dense, external shell; [OIII] emissions arise in a 
relatively smooth, low density interface between the bubble gas and the 
H$\alpha$+[NII] filaments. 

As noticed by Meaburn et al. (1996), in this case 
the [OIII] image is expected to follow the H$\alpha$+[NII] one, and both are 
dominated by the swept-up shell. The alternative explanation proposed by 
Meaburn et al. is based on Mellema's (1995) models: the nebular 
material is swept-up both by the fast wind and by the ionization front. 
So, in the first evolutionary phases, when the central star temperature is 
still relatively low, the two shells appear as distinct features: H$\alpha$ 
and [NII] prevail in the shell swept-up by the H ionization front, and [OIII] 
in the wind-swept shell. In the specific case of NGC 40, the second, faster 
shell has almost undertaken the first one. 

Which model is right? The answer provided by our data is ambiguous: 
on one side we obtain that the 
[OIII] and the H$\alpha$+[NII] emitting regions are disconnected (in many 
cases they are opposite, i.e. strong [OIII] is associated with 
faint H$\alpha$+[NII], 
and vice versa); on the other side the density distribution across the shell 
never shows evidence of a double peaked structure implicit in Mellema's 
models. 

So, for the present, we can only say that: 

- in the 
equatorial region of NGC 40 the stellar wind is blocked by the dense 
nebular gas and [OIII] emissions occur in the innermost, faster expanding 
part of the "bell" density profile; 

- at the poles, where densities are lower than 
at the equator, the braking effect by the nebular gas is less efficient 
and [OIII] appears in extended zones, internally to the polar caps. 

A question 
arises: since we observe modest expansion velocities in the nebular gas, where 
(and how) is decelerated the stellar wind, whose original speed is of about 
2000 km s$^{-1}$?

It is evident that further, more accurate observations are needed not only 
for the nebula, but also for the central star. 
At the moment, mass-loss rates 
and terminal wind velocities reported in the literature for the WC8 nucleus of 
NGC 40 span in the range 3x10$^{-8}$ M$_\odot$ yr$^{-1}$ and 2600 km s$^{-1}$ 
(Cerruti-Sola and Perinotto, 1985) to  1x10$^{-6}$M$_\odot$ yr$^{-1}$ and 
1800 km s$^{-1}$ (Bianchi, 1992). Moreover, the presence of wind fluctuations 
was recently discovered by Balick et al. (1996) and Acker et al. (1997). 

The ionized nebular mass obtained from 
the observed H$\beta$ flux (Sabbadin et al., 1987) results to be M$_{ion}$=
2 - 5 x10$^{-2}$ M$_\odot$ (depending on the adopted values of F(H$\beta$), 
c(H$\beta$), distance, angular radius, electron density and electron 
temperature), and a conservative upper limit to the nebular mass pumped up 
by the fast stellar wind can be put at 1x10$^{-2}$ M$_\odot$. 

Even in the 
most favorable case (i. e. momentum conservation), Bianchi's wind 
(10$^{-6}$ M$_\odot$ yr$^{-1}$ and 1800 km s$^{-1}$) needs less than 20 years 
to produce the observed nebular acceleration (and in this case, where, and 
how, 
is dissipated the wind luminosity, amounting to 500 L$_\odot$, a value which 
is at least ten times the nebular luminosity?). 

A longer interaction time 
(500 years) 
is obtained with the mass-loss rate and wind velocity reported by Cerruti-Sola 
and Perinotto (1985); in this case the wind luminosity is 33 L$_\odot$, of the 
same order of magnitude of the nebular luminosity.  

A final question concerns NGC 40 as prototype of a small group of very low 
excitation planetary nebulae powered by ``late'' WR stars: are the 
peculiarities found in NGC 40 (in particular, the positive inwards velocity 
gradient and the radial chemical composition gradient) a common characteristic
 of these nebulae (e. g. SwSt 1, PM 1-188, Cn 3-1, BD+30$^o$3639, He 2-459,
 IRAS 21282, M 4-18, He 2-99, He 2-113, He 2-142, He 3-1333, Pe 1-7 and K 
2-16)?

Though a definitive answer will need accurate imagery and spectroscopy 
of a suitable sample of objects, a first indication comes from the very recent
results by Bryce and Mellema (1999), showing that BD+30$^o$3639 expands 
faster in [OIII] than in [NII].

\section{Conclusions}

Long-slit echellograms at different position angles of the bright core of 
NGC 40 allowed to 
analyse the expansion velocity field (the higher is the ionization, the faster
is the motion), the radial density distribution (a "bell" profile reaching 
$Ne$=4000
cm$^{-3}$), the radial ionization structure (peculiar in the innermost 
parts), and the
chemical abundances (a gradient is present, due to contamination of the
hydrogen-depleted stellar wind). 

Moreover, tomographic maps are obtained, 
giving the spatial matter distribution along the cross-sections of nebula 
covered
by the slit. 

All these observational results, discussed within the interacting 
winds model, point out our incomplete, ``nebulous'' knowledge
of the physical processes shaping this peculiar planetary nebula where 
``the tail wags the dog'' (as wittily suggested by Prof. Lawrence Aller, 
private communication).

\begin{acknowledgements}
We wish to express our gratitude to Professors Lawrence Aller and Manuel 
Peimbert 
and to Drs. Luciana Bianchi, Francesco Strafella and Vittorio Buond\'\i\ 
for help, suggestions and encouragements.

We thank the whole technical staff -in particular the night 
assistants- of the Astronomical Observatory of Asiago at Cima Ekar for their
competence and patience.

This paper is partially based on observations made with the Italian Telescopio
 Nazionale Galileo (TNG) operated on the island of La Palma by the Centro 
Galileo Galilei of the CNAA (Consorzio Nazionale per l'Astronomia e 
l'Astrofisica) at the Spanish Observatorio del Roque de los Muchachos of the 
Instituto de Astrofisica de Canarias.  
\end{acknowledgements}

\bibliographystyle{astron}
\bibliography{/home/enrico/testi/bibliografia}

\end{document}